# Image Synthesis with Graph Cuts: A Fast Model Proposal Mechanism in Probabilistic Inversion

Tobias Zahner[1], Tobias Lochbühler[1], Grégoire Mariethoz[2,3], Niklas Linde[1]

[1]*Applied and Environmental Geophysics Group, Institute of Earth Sciences, University of Lausanne, 1015 Lausanne, Switzerland. E-mail:* Niklas.Linde@unil.ch

[2]*Institute of Earth Surface Dynamics, University of Lausanne, 1015 Lausanne, Switzerland.*

[3]*School of Civil and Environmental Engineering, UNSW Australia, Sydney NSW 2052, Australia.*

# SUMMARY

Geophysical inversion should ideally produce geologically realistic subsurface models that explain the available data. Multiple-point statistics is a geostatistical approach to construct subsurface models that are consistent with site-specific data, but also display the same type of patterns as those found in a training image. The training image can be seen as a conceptual model of the subsurface, and is used as a non-parametric model of spatial variability. Inversion based on multiple-point statistics is challenging due to high non-linearity and time-consuming geostatistical resimulation steps that are needed to create new model proposals. We propose an entirely new model proposal mechanism for geophysical inversion that is inspired by texture synthesis in computer vision. Instead of resimulating pixels based on higher-order patterns in the training image, we identify a suitable patch of the training image that replace a corresponding patch in the current model without breaking the patterns found in the training image, that is, remaining consistent with the given prior. We consider three cross-hole ground-penetrating radar examples, in which the new model proposal mechanism is employed within an extended Metropolis Markov chain Monte Carlo (MCMC) inversion. The model proposal step is about 40 times faster than state-of-the-art multiple-point statistics resimulation techniques, the number of necessary MCMC steps is lower and the quality of the final model realizations are of similar quality. The model proposal mechanism is presently limited to two-dimensional fields, but the method is general and can be applied to a wide range of subsurface settings and geophysical data types.

# 1 INTRODUCTION

Geophysical inversion can produce subsurface models that are consistent with available geophysical data and any additional information and constraints. In a deterministic framework, constraints are often synonymous with explicit regularization terms that, for instance, penalize deviations from a reference model [i.e. damping, *Marquardt*, 1963], a smooth model [*Constable et al.*, 1987] or a multi-Gaussian geostatistical model [e.g., *Maurer et al.*, 1998]. Regularization operators are known to have a strong impact on the resulting subsurface models [e.g., *Ellis and Oldenburg*, 1994]. Probabilistic Bayesian inversion replaces the regularization constraints by a prior probability density function (pdf) that describes all information that is available other than the geophysical data to be inverted. The probabilistic approach explicitly searches for multiple model realizations that agree with the data and available prior information. Probabilistic inversion is often performed using global methods, such as Markov chain Monte Carlo (MCMC) methods which aim at sampling a posterior pdf by combining the data likelihood (how likely it is that a proposed model gave rise to the observed data) and the prior pdf. In these methods, there is no need to assume a linear or quasi-linear relation between the data and the model parameters, neither must the posterior pdf or the data likelihood be continuous and it might exhibit multiple maxima. If properly constructed, the retrieved samples from the MCMC chain quantify the state of knowledge about the subsurface [e.g., *Sambridge and Mosegaard*, 2002; *Tarantola*, 2005].

An important and largely outstanding question is how to define and formulate a prior pdf that accounts for the expected spatial variability. Classically, an explicit statistical model (e.g., a multi-Gaussian model) is assigned that describes the expected mean value, as well as model parameter variations and their spatial correlations. However, for many geological situations, it can be difficult to formulate this pdf such that it produces model realizations that can be considered realistic, or at least acceptable, by a geologist [*Zinn and Harvey*, 2003]. Since 20 years, multiple-point statistics (MPS) offers means to produce geologically plausible subsurface realizations [e.g., *Guardiano and Srivastava*, 1993; *Strebelle*, 2002; *Hu and Chugonova*, 2008; *Mariethoz and Caers*, 2014]. The underlying idea is to replace the mathematically tractable statistical model with a sampling procedure that draws higher-order patterns using a so-called training image [*Emery and Lantuéjoul*, 2014]. In the field of earth sciences, training images are 2-D or 3-D representations that describe the expected spatial continuity and properties of geological structures [*Caers and Zhang*, 2004]. For example, process-based simulations, expert knowledge, outcrops or even geological sketches can serve as basis for obtaining training images [*Strebelle*, 2002]. The concept is general and multiple-point statistics has been used in many different disciplines, such as, for example, seismic inversion [*Gonzales et al.*, 2008], hydrogeology [*Kessler et al.*, 2013], mining [*Rezaee et al.*, 2014], porous media reconstruction [*Tahmasebi and Sahimi*, 2013], remote sensing [*Ge and Bai*, 2011; *Jha et al.*, 2013], soil science [*Meerschman et al.*, 2013], geomorphology [*Pirot et al.*, 2014; *Vannametee et al.*, 2014] and medical imaging [*Pham*, 2012].

Different multiple-point statistics methods have been developed to produce model realizations that feature the same patterns as those found in the training image [*Arpat and Caers*, 2007; *Hu and Chugunova*, 2008; *Strebelle*, 2002; *Zhang et al.*, 2006]. In these methods, the individual nodes of the model realisations are simulated sequentially, by accounting for local neighbourhood relationships that are present in the training image. At the same time, several inversion techniques specifically designed for being applied in the context of training images have emerged [*Caers*, 2007; *Khaninezhad et al.*, 2012; *Khodabakhshi and Jafarpour*, 2013; *Li et al.*, 2013; *Zhou et al.*, 2012]. In the context of using MCMC with training image based models, it is possible to perturb a given model realization by resimulating only a fraction of the model domain [*Fu and Gomez-Hernandez*, 2009; *Hansen et al.*, 2012; *Mariethoz et al.*, 2010a].

There are a growing number of geophysical studies that incorporate multiple-point statistics concepts [e.g., *Cordua et al.*, 2015; *Cordua et al.*, 2012; *Hansen et al.*, 2012; *Lange et al.*, 2012; *Lochbühler et al.*, 2014; *Lochbühler et al.*, 2015]. Probabilistic MCMC inversion with multiple-point statistics model proposals is promising, but is presently largely hindered by high computational costs. One of these computational bottlenecks is related to the construction of model proposals by time-consuming multiple-point statistics resimulation.

Advances in the field of computer vision have made image synthesis a standard procedure in graphic design. In a recent review, *Mariethoz and Lefebvre* [2014] outlined the similarity between synthesizing surface textures in animation movies and problems occurring in the field of geostatistical simulation. In this context, *Mahmud et al.* [2014] adapted a method known as image quilting [*Efros and Freeman*, 2001] to geostatistical applications and extended it from 2D to 3D. Image quilting is a very fast method to build stochastic realizations, but it is not appropriate to update an existing realization. Indeed, the patches are made up of squares or cubes and the cuts are performed sequentially along an unilateral path [*Daly*, 2004], which makes it difficult to update one patch without breaking the consistency within the entire realization.

We investigate herein the potential of using graph cuts based image synthesis [*Boykov and Kolmogorov*, 2004; *Boykov et al.*, 2001] as a model proposal mechanism in probabilistic geophysical inversion. Similarly to the case of image quilting, the use of graph cuts is inspired by texture synthesis for graphic design applications [*Kwatra et al.*, 2003]. Our objectives are to investigate (1) if graph cut algorithms can be adapted for geophysical inversion, (2) if model proposals based on graph cuts can significantly decrease computing times in training image based probabilistic inversion, and (3) if the quality of the resulting model realizations are comparable or even better than those obtained by state-of-the-art methods.

# 2 METHODOLOGY

## 2.1 The Bayesian formulation of the inverse problem

In Bayesian theory, a vector of model parameters **m** is described in terms of a pdf. Bayes theorem is used to combine the prior information on these uncertain model parameters with the information gained from direct or indirect site-specific data. This combination of information is formalized by Bayes' Theorem:

$$\sigma(\mathbf{m}) = k \cdot L(\mathbf{m})\rho(\mathbf{m}), \tag{1}$$

where $\rho(\mathbf{m})$ and $\sigma(\mathbf{m})$ are the prior and the posterior pdfs, respectively, and $k$ is a normalisation constant that can be ignored if the model parameterization is fixed. The likelihood function $L(\mathbf{m})$ compares the predicted forward response with the measured data for a given description of data and modelling errors. The solution to equation (1) is generally not available in analytical form and it needs to be approximated by sampling from $\sigma(\mathbf{m})$.

Markov chain Monte Carlo (MCMC) methods sample $\sigma(\mathbf{m})$ by performing random walks through the model space [e.g., *Mosegaard and Tarantola, 1995; Sambridge and Mosegaard*, 2002; *Tarantola*, 2005]. Consider a given model realisation $\mathbf{m}_{cur}$ from a *M*-dimensional model space $\mathcal{M}$. The model realisation $\mathbf{m}_{cur}$ can be perturbed by slightly changing the values of all or some of its parameters to create a model proposal $\mathbf{m}_{prop}$. A perturbation that leads from $\mathbf{m}_{cur}$ to $\mathbf{m}_{prop}$ is referred to as a proposal step and a succession of such steps can be seen as a walk (chain) through $\mathcal{M}$.

Each step in a MCMC random walk has two possible outcomes: (i) either a move is made and $\mathbf{m}_{prop}$ becomes the new $\mathbf{m}_{cur}$, or (ii) no move is made and $\mathbf{m}_{cur}$ remains $\mathbf{m}_{cur}$. In the Metropolis sampling algorithm a symmetric proposal distribution is used and a move is made with acceptance probability [*Metropolis et al.*, 1953]:

$$P_{acc} = \min\left(1, \frac{L(\mathbf{m}_{prop})\rho(\mathbf{m}_{prop})}{L(\mathbf{m}_{cur})\rho(\mathbf{m}_{cur})}\right). \qquad (2)$$

When it is impossible to define $\rho(\mathbf{m})$ explicitly, a useful option is to consider outcomes of a geostatistical simulation algorithm as draws from $\rho(\mathbf{m})$. To be efficient it is important to use a proposal distribution that ensures that $\mathbf{m}_{prop}$ is somewhat close to $\mathbf{m}_{cur}$, for example, by only resimulating a fraction of the nodes in the model domain [e.g., *Mariethoz et al.*, 2010a]. The resulting MCMC method is known as extended Metropolis sampling [*Hansen et al.*, 2012; *Mosegaard and Tarantola*, 1995] and $P_{acc}$ is defined as:

$$P_{acc} = \min\left(1, \frac{L(\mathbf{m}_{prop})}{L(\mathbf{m}_{cur})}\right). \qquad (3)$$

## 2.2 Training images as prior information

Subsurface heterogeneity is traditionally quantified by variograms [*Chilès and Delfiner*, 1999; *Goovaerts*, 1997]. Unfortunately, the information contained in the variograms is often insufficient to describe complex geological systems [*Journel and Zhang*, 2006; *Neuweiler et al.*, 2011; *Western et al.*, 1998; *Zinn and Harvey*, 2003]. Multiple-point statistics tools simulate values of individual grid nodes based on multiple points in their neighbourhood, a so-called data event [*Strebelle*, 2002]. This implies that the model of spatial variability is gained from training images.

In this work, we will compare the performance of a new model proposal mechanism that is based on graph cuts with the iterative spatial resampling (ISR) method that uses the direct sampling (DS) multiple-point statistics algorithm as its main building block [*Mariethoz et al.*, 2010a; *Mariethoz et al.*, 2010b]. To create a model perturbation, direct sampling scans the training image until it finds a similar neighbourhood. The value of the point to be simulated is then given by the value at the corresponding point in the training image. At each step in the MCMC chain, the iterative spatial resampling technique keeps a fixed percentage $\varphi$ of the $M$ parameter values in $\mathbf{m}_{cur}$ as conditioning points to simulate a new realisation $\mathbf{m}_{prop}$ with direct sampling. In other words, a fraction $\varphi$ of the parameter values in $\mathbf{m}_{prop}$ remains the same as $\mathbf{m}_{cur}$, while the remaining fraction $(1-\varphi)$ is resimulated using direct sampling.

## 2.3 A new model proposal mechanism based on graph cuts

Texture synthesis techniques are used in graphic design to create new images such as texturized landscapes for animation movies or video games, based on samples called examplars or training images [e.g., *Efros and Freeman*, 2001; *Lasram and Lefebvre*, 2012]. The newly generated images should feature similar textural properties as a training image and they need to be non-repetitive. A popular technique in graphic design to create new images is to form a sort of collage by assembling irregular pieces of the training image [*Kwatra et al.*, 2003]. The shapes of the pieces are adjusted to create transitions that are as seamless as possible, thereby limiting the creation of discontinuities in the simulated image. In this work, we propose to create model perturbations by replacing single patches in $\mathbf{m}_{cur}$ with patches from the training image.

Fig. 1(a) displays a training image (this image is in practice much larger) and an initial model $\mathbf{m}_{cur}$. A random section of the training image, $\mathbf{m}_{ti}$, is chosen that has the same dimension as $\mathbf{m}_{cur}$. A model proposal $\mathbf{m}_{prop}$ is then synthesized from $\mathbf{m}_{cur}$ and $\mathbf{m}_{ti}$. The red dashed line in Fig. 1(b) indicates the optimized cut that defines a patch $\mathbf{m}_{patch}$ of $\mathbf{m}_{ti}$. This $\mathbf{m}_{patch}$ is pasted in $\mathbf{m}_{cur}$ to create $\mathbf{m}_{prop}$.

Panels 1-3 in the lower part of Fig. 1 show how such a cut is found:

1) The difference image $\boldsymbol{\delta} = |\mathbf{m}_{cur} - \mathbf{m}_{ti}|$ quantifies the discrepancy between $\mathbf{m}_{cur}$ and $\mathbf{m}_{ti}$. It is formed by taking the absolute difference of the pixel values (i.e., the values of the geophysical property under consideration) in $\mathbf{m}_{cur}$ and $\mathbf{m}_{ti}$.

2) Two disconnected regions of high difference, $\mathbf{s}$ (blue) and $\mathbf{t}$ (green), of similar size, are randomly selected.

3) The size and shape of the patch $\mathbf{m}_{patch}$ taken from $\mathbf{m}_{ti}$ is given by the trajectory (red dashed line) of a minimum cut [*Boykov and Kolmogorov*, 2004] that separates $\mathbf{s}$ and $\mathbf{t}$.

The next sections describe these steps in details.

## 2.4 Principle of min-cut/max-flow algorithms

A digital image can be described as a graph or network $\mathbf{G} = [\boldsymbol{\delta}, \mathbf{e}]$ that consists of pixels or nodes $\delta \in \boldsymbol{\delta}$ with fixed relative positions that are represented as connectors (edges $e \in \mathbf{e}$) between nodes (the cut indicated by a red dashed line in Fig. 1 is a cut along edges). Our focus is on unidirected graphs, where each pair of connected nodes $\delta_j$ and $\delta_k$ has a single edge $e_{jk} = (\delta_j, \delta_k) = (\delta_k, \delta_j)$ *[Boykov and Funka-Lea, 2006]*.

In graph theory, a cut partitions a graph into two disjoint subsets. Subsets are disjoint when they do not share any elements (nodes or edges) [*Ford and Fulkerson*, 1962]. The difference image $\boldsymbol{\delta}$ in Fig. 1 is an example of such a graph. To explain the basic theory, we consider first a very simple graph $\mathbf{G}$ with four nodes $\delta_1, \delta_2, \delta_3, \delta_4$ and four edges $e_{1,2}$, $e_{1,3}$, $e_{2,4}$, $e_{3,4}$ (Fig. 2a). Fig. 2(e) shows one out of six possible partitions of graph $\mathbf{G}$ into two disjoint subsets $\mathbf{S}$ and $\mathbf{T}$ by a cut $\mathbf{C}$ going along edges $e_{1,3}$ and $e_{2,4}$. An efficient method to find a suitable cut is to consider the graph as a network of pipes with a corresponding flow through the graph [*Boykov and Kolmogorov*, 2004; *Greig et al.*, 1989]. The process of finding a preferential cut is here described by considering graph $\mathbf{G}$, in which each edge has been assigned a capacity: $c_{1,2} = 2, c_{1,3} = 2, c_{2,4} = 1$ and $c_{3,4} = 3$. We assign the capacities of the edges/connectors as the sum of the difference values of the two pixels they connect (see inset in Fig. 1, panel 3). This formulation of capacity is not physically consistent with the capacity of two connected pipes, but it is widely used in the compiuter graphics community [*Boykov and Kolmogorov*, 2004] to ensure that the "flow" is high in regions of high difference values. The cost to cut an edge is equal to the capacity of the edge, which implies that the total cost of a cut is the sum of the capacities of the edges along its path. The problem of finding the best cut can then be formulated as an energy minimisation problem, where water flows through a network from a source to a sink region. The nodes of the source $\mathbf{s} = [s_1, s_2, s_3, ...]$ and the sink $\mathbf{t} = [t_1, t_2, t_3, ...]$

regions are internally connected by edges/connectors of unlimited capacity. This is done to avoid cuts through the source or sink regions.

For graph **G**, the source $\mathbf{s} = [\delta_1]$ and sink $\mathbf{t} = [\delta_2]$ regions consist each of only one node (see Fig. 2c). At maximum steady state flow from source to sink, the edges $e_{1,3}$ and $e_{2,4}$ are at their capacity limits ($c_{1,3} = 2, c_{2,4} = 1$) as indicated by the red colour in Fig. 2(d). This implies that it is these two edges that control the maximum flow that can transit through the graph. Fig. 2(e) shows a cut along edges (red dotted line) that separates **s** and **t** by creating two disjoint subsets **S** (in blue) and **T** (in green). It has a cost $C = \sum c_{cut} = c_{1,3} + c_{2,4} = 2 + 1 = 3$ which is equal to the maximal steady state flow through the network. Such a cut is a minimum cut according to the min-cut/max-flow theorem by *Ford and Fulkerson* [1962]. This theorem might be understood intuitively by looking on Fig. 2(d). The edges highlighted in red are the bottlenecks of the flow system. A minimum cut follows these bottlenecks. To compute the minimum cut, we rely on *Boykov and Kolmogorov* [2004] that developed an optimised open-source min-cut/max-flow algorithm that compares favourably against alternative methods.

## 2.5 Cutting a patch from the training image

To create a model proposal, we determine an appropriate patch by separating the graph $\boldsymbol{\delta} = |\mathbf{m}_{cur} - \mathbf{m}_{ti}|$ into two independent subsets using the principles introduced in section 2.4. Our implementation can be summarized as follows:

(i) Randomly choose a part $\mathbf{m}_{ti}$ of the training image that has the same shape and size as $\mathbf{m}_{cur}$.

(ii) Form the difference image $\boldsymbol{\delta} = |\mathbf{m}_{cur} - \mathbf{m}_{ti}|$.

(iii) Create an outer frame of one pixel thickness around **δ** with node values of $\min(\boldsymbol{\delta})$ (indicated as a dark grey frame in panel 1 in Fig. 1).

(iv) Create a second outer frame of one pixel thickness around **δ** with high node values (e.g., 10 times $\max(\boldsymbol{\delta})$) (indicated as a white frame surrounding the dotted region in panel 1 in Fig. 1).

(v) Find connected components (i.e., subsets of the graph) of node values $\boldsymbol{\delta} \geq \text{mean}(\boldsymbol{\delta})$ by a connected component analysis [*Haralick and Shapiro*, 1992; *Renard and Allard*, 2013].

(vi) If the number of connected components is less than two, then assign the patch **m**$_{\text{patch}}$= **m**$_{\text{ti}}$ and go to step (xi), else:

(vii) Compute the capacities connecting any two neighbouring pixels $\delta_j$ and $\delta_k$ as the sum of their node values $c_i(\delta_j, \delta_k) = \delta_j + \delta_k$ (see Fig. 1). Randomly select a connected component with an area of at least *p* pixels (10 was used in this work) and define it as terminal **s**. Among the remaning connected components, find the one with the most similar area as **s** and assign it as **t**.

(viii) Find the min-cut that separates **s** and **t** [*Boykov and Kolmogorov*, 2004] into two disjoint subsets.

(ix) Choose the smaller of the two subsets as the patch **m**$_{\text{patch}}$.

(x) Paste **m**$_{\text{patch}}$ to **m**$_{\text{cur}}$ to create **m**$_{\text{prop}}$.

The perturbed model realisation **m**$_{\text{prop}}$ is evaluated using the acceptance criterion in eq. (3) and is used to perform one step in the MCMC chain. Min-cut/max-flow algorithms find the best possible update patch within the constraints given by **s**, **t** and the frames around **δ**. We have found that defining **s** and **t** as irregularly shaped areas based on the difference image (step v above) provides fewer artifacts than assigning them randomly using a regular shape, such as a rectangle of a pre-defined size. The two frames around $\boldsymbol{\delta}$ force the cut to form a closed path. Without the frame, the cuts tend to form paths that connect two points at the border of $\boldsymbol{\delta}$. This leads to excessively large model updates, which decreases the efficiency of the MCMC algorithm.

# 3 TEST EXAMPLES

Our new method has been tested on synthetic cross-hole ground penetrating radar (GPR) data using three different training images. Test cases I and II consider binary images with two facies of homogeneous GPR velocity (60 m/μs and 80 m/μs). Test case I (Fig. 3a) resembles river channels, while test case II, (Fig. 4a) resembles a system with sand lenses. Test case III (Fig. 5a) considers a continuous example in the form of a multi-Gaussian random field.

## 3.1 Model setup

The crosshole GPR method uses a transmitter antenna to emit a high-frequency electromagnetic wave in one borehole and a receiver antenna to record the arriving energy in another borehole [e.g., *Peterson*, 2001; *Annan*, 2005]). We consider first-arrival travel times for various transmitter and receiver locations. These data provide constraints on the GPR velocity distribution between the boreholes. The GPR velocity is primarily a function of electric permittivity, which is strongly dependent on the water content and, hence, porosity in saturated media.

We consider two vertical boreholes that are located 5.0 m apart. Sources (left) and receivers (right) are placed between 0.5 and 10.5 m depth with 0.4 m spacing. Only source-receiver combinations with an angle of less than ±50° to the horizontal are considered [*Peterson*, 2001], leading to a total dataset of $N = 566$ travel times.

Fig 3(a) shows the true GPR velocity field $\mathbf{m}_{true}$ for test case I. Synthetic travel time data **d**, were simulated by solving the forward problem for $\mathbf{m}_{true}$ and adding an error:

$$\mathbf{d} = g(\mathbf{m}_{true}) + \boldsymbol{\varepsilon}, \qquad (4)$$

where $\boldsymbol{\varepsilon}$ represents independent random draws from a Gaussian distribution with mean $\mu=0$ ns and a typical standard deviation $\sigma=1$ ns. The first-arrival travel times are computed using the Eikonal equation solver by *Podvin and Lecomte* [1991]. The data used for the other two test cases (Figs 4a and 5a) were generated in an analogous manner. Under the given Gaussian error model, the likelihood function is [*Tarantola*, 2005]:

$$L(\mathbf{m}) = \left(\frac{1}{\sqrt{2\pi\sigma^2}}\right)^N \exp\left[-\frac{1}{2}\left((\mathbf{d}-g(\mathbf{m}))^T \mathbf{C}_d^{-1} (\mathbf{d}-g(\mathbf{m}))\right)\right] \qquad (5)$$

with the data covariance matrix $\mathbf{C}_d$ being a diagonal matrix with entries $\sigma^2$.

We use the model perturbation procedure described in section 2.3 within an extended Metropolis MCMC sampling scheme. A training image representing an area of 250 × 250 meters with a discretization of 0.1 m was used. We stress here that the reference field $\mathbf{m}_{ref}$ used to create the synthetic data is not part of the training image itself. The rather large size of the training image was chosen to avoid underestimating the variability of the prior. Fig. 3(b) displays a section of the training image that was used for test case I.

## 3.2 Results

To visualize our prior, we run one initial chain of 100'000 steps in which we accept every $\mathbf{m}_{prop}$ (see Mosegaard and Tarantola, 1995). Four random model realizations from $p(\mathbf{m})$ are shown in Fig. 3(c), while Figs 3(d)-(e) show its mean $\mathbf{m}_\mu$ and standard deviation $\mathbf{m}_\sigma$. There are no obvious artifacts in these results that could be attributed to the graph cut algorithm and the prior model realizations look similar to random sections of the training image. The inferred values of $\mathbf{m}_\mu$ and $\mathbf{m}_\sigma$ are close to the mean and standard deviation of the training image, respectively (see Table 1). Small-scale variations in $\mathbf{m}_\sigma$ (see Fig. 3e) can be attributed to the finite size of the set of prior model realizations.

To infer $\sigma(\mathbf{m})$, we consider for each test example the results of five independent MCMC chains of one million steps. Four random model realisations from $\sigma(\mathbf{m})$ are shown in Fig 3(f), while Figs 3(g)-(h) show its mean $\mathbf{m}_\mu$ and standard deviation $\mathbf{m}_\sigma$. The individual model realizations resemble $\mathbf{m}_{true}$ (Fig. 3a) closely. The blurriness of $\mathbf{m}_\mu$ (Fig. 3g) is attributed to model uncertainty that is quantified by the standard deviation $\mathbf{m}_\sigma$ (Fig. 3h). It is seen that the uncertainty is the highest around areas with sharp discontinuities.

Figs 4 and 5 show corresponding results for test cases II and III, respectively. The results for test case I and II are fairly similar in nature. The continuous test case III shows more variability in the posterior model realizations compared to the other test cases (Fig. 5f).

## 3.3 Convergence analysis

A MCMC chain needs a burn-in time before it reaches regions of high posterior probability and starts to sample the posterior pdf. Accordingly, the model realisations sampled during the burn-in phase are not considered part of the posterior pdf. Herein, we define the burn-in period as the steps preceeding the point where WRMSE ≤ 1 is reached for the first time. The weighted root mean square error (WRMSE) is a measure that is related to the logarithm of the likelihood function:

$$WRMSE = \sqrt{\frac{1}{N} \sum_{i=1}^{N} \frac{(d_i - g_i(\mathbf{m}))^2}{\sigma^2}} \ . \tag{6}$$

This measure of data misfit is often used in classical deterministic inversions and a value of 1 indicates that the data residuals have similar magnitudes as the standard deviation of the data errors [e.g., *Constable et al.*, 1987].

The WRMSE is shown for all five MCMC chains for each test case (Fig. 6a). The burn-in time is approximately 200'000 steps for test case I, 150'000 steps for test case II and only 5'000 steps for test case III. For reasons of simplicity and to be on the safe side, model realisations of the first 200'000 steps are not considered further for any of the test cases.

The mean of the autocorrelation coefficients (acf) provides a measure of the within-chain correlation after burn-in (Fig. 6b). A value of one indicates complete correlation and a value of zero none. As expected, the average correlation decreases with increasing lags of MCMC steps. The acceptance rate is on average 5 % for test case I, 3 % for test case II and 20 % for test case III. The distribution of the model fraction that is in each model proposal step after burn-in is illustrated in Fig. 6(c). In particular for test case II, it regularly happens that there is only one connected component of the difference image. No graph cut is then possible and the model proposal is the complete patch of the training image (see step V in section 2.5). After burn-in, it is most unlikely that such proposals with a model fraction of 1 will be accepted.

The path taken to sample the posterior model realizations depends on the starting point. One common way to assess if the chains reach the same limiting distribution is to compare the outcome of several chains that start at different points in the model space. The potential scale reduction factor $\hat{R}$ [*Gelman and Rubin*, 1992] does so by comparing the average within-chain variance with the variance of the within chain means for the second half of the chain. For each test case, we calculate $\hat{R}$ values for all model parameters. Fig. 7 shows the corresponding $\hat{R}$ values for 5 different chain lengths *2n* from 2'000 steps to 1'000'000. As expected, the $\hat{R}$ values decrease with the number of steps. The rate of the decrease differs significantly between parameters and test cases. Convergence of the MCMC chain is typical declared when $\hat{R} \leq 1.2$ for all individual pixels. When comparing the results for test case I (Fig. 7a) with $\mathbf{m}_{true}$ (see Fig 3a), it can be noticed that values decrease slower around areas with sharp discontinuities. The same observation holds for test case II (Fig. 7b). Compared to Figs. 7(a)-(b), the $\hat{R}$ values for test case III (Fig. 7c) generally decrease quicker.

To quantify the similarity between prior realisations and the training image, we compare the experimental variograms of the training image with those of the sampled prior (Fig. 8a). We considered 100 random prior realizations and 100 sections of the training image. The results indicate that the average range and sill of the variograms are similar. This also holds for test cases II (Fig. 8b) and III (Fig. 8c).

The experimental variograms of 100 randomly chosen posterior model realizations are now compared with the variogram of $\mathbf{m}_{true}$ (Fig. 8d). The variability between the variograms is much smaller for the posterior samples than for the prior samples (Fig. 8a). Similar results are obtained for test cases II (Fig. 8e) and III (Fig. 8f). However, for test case III the semivariance is slightly underestimated up to a lag of 0.3 m, and slightly overestimated for larger lags when compared with the variogram of $\mathbf{m}_{true}$.

### 3.4 Comparison with iterative spatial resampling

For test case I, we now compare our results by replacing our model perturbation step with the one used in iterative spatial resampling (see section 2.2). Our tests on a standard personal computer without any parellization step indicate that model perturbation by the graph-cut-based method is about 38 times faster than with direct sampling (an average time of 0.0944 vs 3.5948 s per model proposal). A fourfold increase in the number of pixels lead to a 27% longer computational time for the graph cut, while the direct sampling algorithm needed 600% longer computational time. This suggests that the computational efficiency of the graph cut algorithm becomes even more pronounced in higher dimensions. The comparison was made with both perturbation methods implemented in Matlab, but with the core of the code (i.e., the min-cut/max-flow or the direct sampling algorithm) in optimized C. Due to the higher computational effort of direct sampling, we limited the MCMC chains to 30'000 steps only. Fig. 9(a) compares the burn-in phase for both methods.

*Mariethoz et al.* [2010a] showed that choosing an appropriate update size $\varphi$ is important to reach the target misfit. In our examples we found that using a constant value of $\varphi = 10\%$ lead to the quickest convergence compared to other constant values. Nevertheless, convergence towards the target WRMSE was further improved by varying $\varphi$ randomly between 0 and 20% according to a uniform distribution. The convergence of the WRMSE is slower with direct sampling than with graph cuts. This indicates that a longer burn-in phase than 200'000 steps would be needed. For this example, it was impossible to sample meaningful posterior realizations using only 30'000 steps.

Increasing the data errors or using fewer data makes it much easier to reach the target misfit and fewer MCMC steps are needed for convergence. *Ruggeri et al.* (2015) present a detailed evaluation of these effects in the context of multi-Gaussian fields and crosshole GPR data; we refer the interested reader to this publication. To compare both methods within our computational budget, we repeated the inversion using $N$=29 travel times only (to do so we changed the source and receiver spacing from 0.4 to 2.0 m). For this case, the convergence is faster and the posterior uncertainty is larger. Fig. 9(b) is analogous to Fig. 9(a) for the sparse-data case. The burn-in phase decreased to less than 6'000 steps for the chains that use graph cuts or direct sampling with a randomly varying $\varphi$. Fig 10(a) shows four random posterior model realisations created with graph cuts, while Figs 10(b)-(c) show the corresponding $\mathbf{m}_\mu$ and $\mathbf{m}_\sigma$. Four randomly chosen posterior model realisations are shown in Fig. 10(d), while Figs 10(e)-(f) show the corresponding $\mathbf{m}_\mu$ and $\mathbf{m}_\sigma$.

## 4 DISCUSSION

While all test cases presented in this paper are 2-dimensional, nothing prevents extending the approach to 3D cases. The graph formulation remains identical regardless of the dimensionality of the images considered. This is a feature mentioned in the initial graph cuts paper [*Kwatra et al.*, 2003] that demonstrates its application on video textures, which are 3D objects. Similarly, the connected components analysis used to identify the terminals of the min-cut/max-flow problem can be carried out in any dimensionality.

One limitation of the approach presented in this paper is the assumption that the simulated field is stationary. This is a consequence of the fact that the new patch is randomly selected from any location in the training image. In multiple-point geostatistics, several approaches have been developed to deal with non-stationary training images, which could potentially be adapted to the context of graph cuts inversion. In particular, the concept of using an auxiliary variable describing the non-stationarity [*Chugunova and Hu*, 2008] could be used to guide the selection of new patches such that non-stationarity constraints are respected.

When using the graph cuts approach, one needs to extract from the training image many different patches having the size of the modelled area. One requirement of the approach is therefore that the training image is much larger than the modelled domain. Using large training images is a general requirement of multiple-point geostatistics, to the point that *Emery and Lantuéjoul* [2014] argued that to present sufficient repetition, the training image should be impractically large. One alternative to using such large training images is to have a mechanism that enriches the generated models with new patterns that are not present in the original training image. This is achieved herein by cutting patches, since the cut can create new patterns that do not exist in the training image. In addition to relaxing the requirement for a large training image, the new patterns may also improve the sampling in a MCMC process by allowing a slight departure from the often too narrow pool of prior patterns present in the training image.

# 5 CONCLUSIONS

We have introduced a new model proposal mechanism in MCMC inversion that is based on image synthesis with graph cuts, and we suggest that it opens up training image based inversion to areas where it has up to now been impractical. This is demonstrated by the presented test cases, in which graph-cut-based image synthesis is used as a proposal mechanism to constrain a solution space on the basis of a training image. By rearranging patches of a training image in new ways, realisations from a prior distribution can be more quickly generated than by using standard multiple-point statistics tools, such as direct sampling. The new proposal mechanism is much faster than classical multiple-point statistics resimulation methods, convergence is at least as fast and the quality of the sampled posterior model realizations is comparable. Our method is demonstrated for cross-hole GPR data but the general-purpose concept can be applied to a wide range of geophysical and other geoscientific (e.g., hydrogeological) data. Even if the test cases are somewhat simplistic, we find that the method is versatile as it works well for both continuous and discretely varying fields. Moreover, there are no free algorithmic parameters that must be tuned in the inversion. In future work we plan to test the graph cuts method on field data using more advanced training images.

# ACKNOWLEDGEMENTS

This research was funded by the Swiss National Science Foundation (SNF) and is a contribution to the ENSEMBLE project (grant no. CRSI22_132249). We would like to thank Gianluca Fiandaca and an anonymous reviewer for their constructive and detailed comments that helped to significantly improve this paper. The first, third and fourth authors dedicate this work to the memory of Tobias Lochbühler who tragically lost his life in a mountaineering accident on July 19, 2014. Tobias was a truly wonderful person and a most gifted researcher.

# REFERENCES

Annan, A. P. (2005), GPR methods for hydrogeological studies, in Hydrogeophysics, edited by Y. Rubin and S. S. Hubbard, Springer, The Netherlands, pp. 185-214.

Arpat, B., and J. Caers (2007), Conditional simulations with patterns, *Math. Geol.*, *39*(2), 177-203.

Boykov, Y., and G. Funka-Lea (2006), Graph cuts and efficient N-D image segmentation, *International Journal of Computer Vision*, *70*(2), 109-131.

Boykov, Y., and V. Kolmogorov (2004), An experimental comparison of min-cut/max-flow algorithms for energy minimization in vision, *Pattern Analysis and Machine Intelligence, IEEE Transactions on*, *26*(9), 1124-1137, doi:10.1109/TPAMI.2004.60.

Boykov, Y., O. Veksler, and R. Zabih (2001), Fast approximate energy minimization via graph cuts, *IEEE Transactions on Pattern Analysis and Machine Intelligence*, *23*(11), 1222-1239.

Caers, J. (2007), Comparing the gradual deformation with the probability perturbation method for solving inverse problems, *Math. Geol.*, *39*(1), 27-52.

Caers, J., and T. Zhang (2004), Multiple-point geostatistics: a quantitative vehicle for integrating geologic analogs into multiple reservoir models, in *Integration of outcrop and modern analog data in reservoir models, AAPG memoir 80*, edited by G. M. Grammer, Harris, P. M., and Eberli, G. P., pp. 383-394, American Association of Petroleum Geologists, Tulsa.

Chilès, J.-P., and P. Delfiner (1999), *Geostatistics - Modeling Spatial Uncertainty*, 695 pp., John Wiley & Sons, Inc., New York, USA.

Chugunova, T., and L. Hu (2008), Multiple-point simulations constrained by continuous auxiliary data, *Math. Geosci.*, *40*(2), 133-146.

Constable, S. C., R. L. Parker, and C. G. Constable (1987), Occam's inversion; a practical algorithm for generating smooth models from electromagnetic sounding data, *Geophysics*, *52*(3), 289-300, doi:10.1190/1.1442303.

Cordua, K., T. Hansen, and K. Mosegaard (2015), Improving the pattern reproducibility of multiple-point-based prior models using frequency matching, *Math. Geosci.*, *47*(3), 317-343, doi:10.1007/s11004-014-9531-4.

Cordua, K. S., T. M. Hansen, and K. Mosegaard (2012), Monte Carlo full-waveform inversion of crosshole GPR data using multiple-point geostatistical a priori information, *Geophysics*, *77*(2), H19-H31.

Daly, C. (2004), Higher order models using entropy, Markov random fields and sequential simulation, paper presented at Geostatistics Banff 2004, Kluwer Academic Publisher, Banff, Alberta.

Efros, A. A., and W. T. Freeman (2001), Image quilting for texture synthesis and transfer paper presented at Proceedings of the ACM SIGGRAPH Conference on Computer Graphics.

Ellis, R. G., and D. W. Oldenburg (1994), Applied geophysical inversion, *Geophysical Journal International*, *116*(1), 5-11, doi:10.1111/j.1365-246X.1994.tb02122.x.

Emery, X., and C. Lantuéjoul (2014), Can a training image be a substitute for a random field model?, *Math. Geosci.*, *46*(2), 133-147, doi:10.1007/s11004-013-9492-z.

Ford, L., and D. Fulkerson (1962), *Flows in networks*, Princeton University Press, Princeton

Fu, J., and J. Gomez-Hernandez (2009), Uncertainty assessment and data worth in groundwater flow and mass transport modeling using a blocking Markov chain Monte Carlo method, *J. Hydrol.*, *2009*(364), 328-341.

Ge, Y., and H. Bai (2011), Multiple-point simulation-based method for extraction of objects with spatial structure from remotely sensed imagery, *International Journal of Remote Sensing*, *32*(8), 2311-2335, doi:10.1080/01431161003698278.

Gelman, A., and D. B. Rubin (1992), Inference from iterative simulation using multiple sequences, *Statistical Science*, *7*(4), 457-472, doi:10.2307/2246093.

Gurdiano, F., and M. Srivastava (1993), Multivariate geostatistics: Beyond bivariate moments. In Soares, A. (ed.) *Geostatistics-Troia*, Kluwer Academic, Dordrecht.

Gonzales, E., T. Mukerji, and G. Mavko (2008), Seismic inversion combining rock physics and multiple-point geostatistics, *Geophysics*, *73*(1), R11-R21, doi:10.1190/1.2803748.

Goovaerts, P. (1997), *Geostatistics for Natural Resources evaluation*, 496 pp., Oxford University Press, Oxford.

Greig, D. M., B. T. Porteous, and A. H. Seheult (1989), Exact maximum a posteriori estimation for binary images, *Journal of the Royal Statistical Society. Series B (Methodological)*, *51*(2), 271-279, doi:10.2307/2345609.

Hansen, T. M., K. S. Cordua, and K. Mosegaard (2012), Inverse problems with non-trivial priors: Efficient solution through sequential Gibbs sampling, *Comput. Geosc.*, *16*(3), 593-611.

Haralick, R. M., and L. G. Shapiro (1992), *Computer and Robot Vision, Volume I*, Addison-Wesley, Boston.

Hu, L., and T. Chugunova (2008), Multiple-Point Geostatistics for Modeling Subsurface Heterogeneity: a Comprehensive Review, *Water Resour. Res.*, *44*(W11413), doi:doi:10.1029/2008WR006993.

Jha, S. K., G. Mariethoz, J. P. Evans, and M. F. McCabe (2013), Demonstration of a geostatistical approach to physically consistent downscaling of climate modeling simulations, *Water Resour. Res.*, *49*(1), 245-259, doi:10.1029/2012WR012602.

Journel, A., and T. Zhang (2006), The necessity of a multiple-point prior model, *Math. Geol.*, *38*(5), 591-610.

Kessler, T. C., A. Comunian, F. Oriani, P. Renard, B. Nilsson, K. E. Klint, and P. L. Bjerg (2013), Modeling fine-scale geological heterogeneity-examples of sand lenses in tills, *Groundwater*, *51*(5), 692-705.

Khaninezhad, M. M., B. Jafarpour, and L. Li (2012), Sparse geologic dictionaries for subsurface flow model calibration: Part I. Inversion formulation, *Adv. Water Resour.*, *39*, 106-121.

Khodabakhshi, M., and B. Jafarpour (2013), A Bayesian mixture-modeling approach for flow-conditioned multiple-point statistical facies simulation from uncertain training images, *Water Resour. Res.*, *49*(1), 328-342.

Kwatra, N., A. Schödl, I. Essa, G. Turk, and A. Bobick (2003), Graphcut textures: Image and video synthesis using graph cuts, *ACM Trans. Graph.*, *22*(3), 277-286.

Lange, K., J. Frydendall, K. S. Cordua, T. M. Hansen, Y. Melnikova, and K. Mosegaard (2012), A frequency matching method: Solving inverse problems by use of geologically realistic prior information, *Math. Geosci.*, *44*(7), 783-803.

Lasram, A., and S. Lefebvre (2012), Parallel patch-based texture synthesis, paper presented at High Performance Graphics, Paris, June 25-27, 2012.

Li, L., S. Srinivasan, H. Zhou, and J. J. Gómez-Hernández (2013), A pilot point guided pattern matching approach to integrate dynamic data into geological modeling, *Adv. Water Resour.*, *62, Part A*(0), 125-138, doi:http://dx.doi.org/10.1016/j.advwatres.2013.10.008.

Lochbühler, T., G. Pirot, J. Straubhaar, and N. Linde (2014), Conditioning of multiple-point statistics facies simulations to tomographic images, *Math. Geosci.*, *46*(5), 625-645, doi:10.1007/s11004-013-9484-z.

Lochbühler, T., J. A. Vrugt, M. Sadegh, and N. Linde (2015), Summary statistics from training images as prior information in probabilistic inversion, *Geophysical Journal International*, *201*(1), 157-171, doi:10.1093/gji/ggv008.

Mahmud, K., G. Mariethoz, J. Caers, P. Tahmasebi, and A. Baker (2014), Simulation of Earth textures by conditional image quilting, *Water Resour. Res.*, *50*(4), 3088-3107, doi:10.1002/2013wr015069.

Mariethoz, G., and J. Caers (2014), *Multiple-Point Geostatistics: stochastic modeling with training images*, 384 pp., Wiley-Blackwell.

Mariethoz, G., and S. Lefebvre (2014), Bridges between multiple-point geostatistics and texture synthesis, *Math. Geosci.*, *66*(66-80), doi:10.1016/j.cageo.2014.01.001.

Mariethoz, G., P. Renard, and J. Caers (2010a), Bayesian inverse problem and optimization with iterative spatial resampling, *Water Resour. Res.*, *46*(11), doi:10.1029/2010WR009274.

Mariethoz, G., P. Renard, and J. Straubhaar (2010b), The Direct Sampling method to perform multiple-point geostatistical simulations, *Water Resour. Res.*, *46*(W11536), 10.1029/2008WR007621.

Marquardt, D. (1963), An algorithm for least-squares estimation of nonlinear parameters, *SIAM Journal on Applied Mathematics*, *11*, 431-441.

Maurer, H., K. Holliger, and D. E. Boerner (1998), Stochastic regularization: Smoothness or similarity?, *Geophys. Res. Lett.*, *25*(15), 2889-2892, doi:10.1029/98GL02183.

Meerschman, E., M. Van Meirvenne, E. Van De Vijver, P. De Smedt, M. M. Islam, and T. Saey (2013), Mapping complex soil patterns with multiple-point geostatistics, *European Journal of Soil Science*, *64*(2), 183-191.

Metropolis, N., A. Rosenbluth, M. Rosenbluth, and A. Teller (1953), Equation of state calculations by fast computing machines, *J. Chem. Phys.*, *21*, 1087-1092.

Mosegaard, K., and A. Tarantola (1995), Monte Carlo sampling of solutions to inverse problems, *J. Geophys. Res.*, *100*(B7), 12,431-412,447.

Neuweiler, I., A. Papafotiou, H. Class, and R. Helmig (2011), Estimation of effective parameters for a two-phase flow problem in non-Gaussian heterogeneous porous media, *J. Contam. Hydrol.*, *120-121*(C), 141-156.

Peterson, J. E. Jr. (2001), Pre-inversion processing and analysis of tomographic radar data. *J. Environ. Eng. Geophys.*, 6, 1-18.

Pirot, G., J. Straubhaar, and P. Renard (2014), Simulation of braided river elevation model time series with multiple-point statistics, *Geomorphology*, *214*(0), 148-156, doi:http://dx.doi.org/10.1016/j.geomorph.2014.01.022.

Podvin, P., and I. Lecomte (1991), Finite difference computation of traveltimes in very contrasted velocity models: a massively parallel approach and its associated tools, *Geophysical Journal International*, *105*(1), 271-284, doi:10.1111/j.1365-246X.1991.tb03461.x.

Renard, P., and D. Allard (2013), Connectivity metrics for subsurface flow and transport, *Adv. Water Resour.*, *51*, 168-196.

Rezaee, H., O. Asghari, M. Koneshloo, and J. Ortiz (2014), Multiple-point geostatistical simulation of dykes: application at Sungun porphyry copper system, Iran, *Stochastic Environmental Research and Risk Assessment*, *28*, 1913-1927, doi:10.1007/s00477-014-0857-8.

Ruggeri, P., J. Irving, and K. Holliger (2015), Systematic evaluation of sequential geostatistical resampling within MCMC for posterior sampling of near-surface geophysical inverse problems. *Geophysical Journal International*, *202*, 961-975.

Sambridge, M., and K. Mosegaard (2002), Monte-Carlo methods in geophysical inverse problems, *Reviews of Geophysics*, *40*(3), 1009, doi:10.1029/2000RG000089.

Strebelle, S. (2002), Conditional simulation of complex geological structures using multiple-point statistics, *Math. Geol.*, *34*(1), 1-22.

Tahmasebi, P., and M. Sahimi (2013), Cross-correlation function for accurate reconstruction of heterogeneous media, *Phys. Rev. Lett.*, *110*(078002).

Tang, Y., P. M. Atkinson, and J. Zhang (2015), Downscaling remotely sensed imagery using area-to-point cokriging and multiple-point geostatistical simulation, *ISPRS Journal of Photogrammetry and Remote Sensing*, *101*(0), 174-185, doi:http://dx.doi.org/10.1016/j.isprsjprs.2014.12.016.

Tarantola, A. (2005), *Inverse Problem Theory and Methods for Parameter estimation*, Society for Industrial and Applied Mathematics, Philadelphia.

Vannametee, E., L. V. Babel, M. R. Hendriks, J. Schuur, S. M. de Jong, M. F. P. Bierkens, and D. Karssenberg (2014), Semi-automated mapping of landforms using multiple point geostatistics, *Geomorphology*, *221*, 298-319, doi:10.1016/j.geomorph.2014.05.032.

Western, A. W., G. Blöschl, and R. B. Grayson (1998), How well do indicator variograms capture the spatial connectivity of soil moisture?, *Hydrological Processes*, *12*(12), 1851-1868.

Zhang, T., P. Switzer, and A. Journel (2006), Filter-based classification of training image patterns for spatial simulation, *Math. Geol.*, *38*(1), 63-80.

Zhou, H., J. Gomez-Hernandez, and L. Li (2012), A pattern search based inverse method, *Water Resour. Res.*, *48*(2), W03505, doi:doi:10.1029/2011WR011195.

Zinn, B., and C. Harvey (2003), When good statistical models of aquifer heterogeneity go bad: A comparison of flow, dispersion, and mass transfer in connected and multivariate Gaussian hydraulic conductivity fields, *Water Resour. Res.*, *39*(3), WR001146.

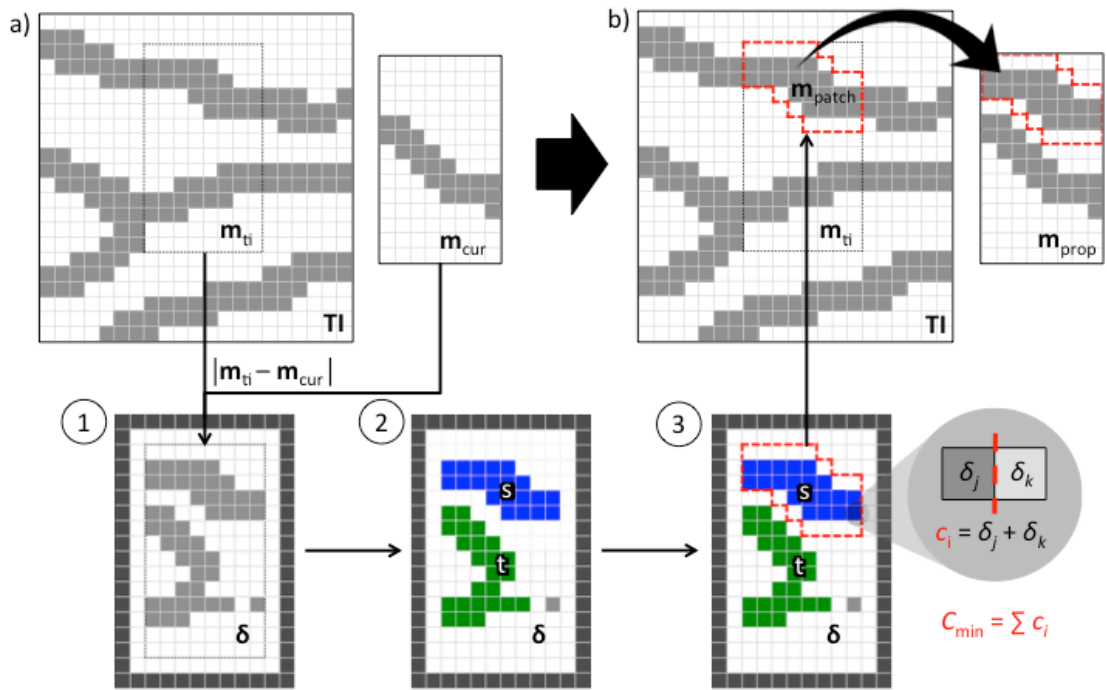

**Figure 1.** Schematic representation of our model proposal mechanism based on graph cuts. (a) A patch of a current model realisation $\mathbf{m}_{cur}$ is (b) replaced with a patch of a random section $\mathbf{m}_{ti}$ of the training image (TI) to form a model proposal $\mathbf{m}_{prop}$ in an MCMC chain. Panels 1 to 3 illustrate how the size and form of the replaced patch is determined: (1) The difference image $\boldsymbol{\delta}$ represents the absolute differences between $\mathbf{m}_{ti}$ and $\mathbf{m}_{cur}$ (high differences are indicated in grey and zero differences are indicated by white). It is framed by a line of nodes with the minimum value of $\boldsymbol{\delta}$ (white), followed by another frame of very high difference (dark grey). (2) Within $\boldsymbol{\delta}$, two terminals $\mathbf{s}$ and $\mathbf{t}$ are defined as randomly chosen areas of high difference of approximately the same size. (3) The piece to be replaced $\mathbf{m}_{patch}$ is given by the cut of minimum cost $C_{min}$ (indicated by a dashed red line) separating $\mathbf{s}$ and $\mathbf{t}$. The capacity $c$ of any cut is given as the sum of the differences of the nodes it is separating. The total cost $C_{min}$ is the sum of these capacities along the cut.

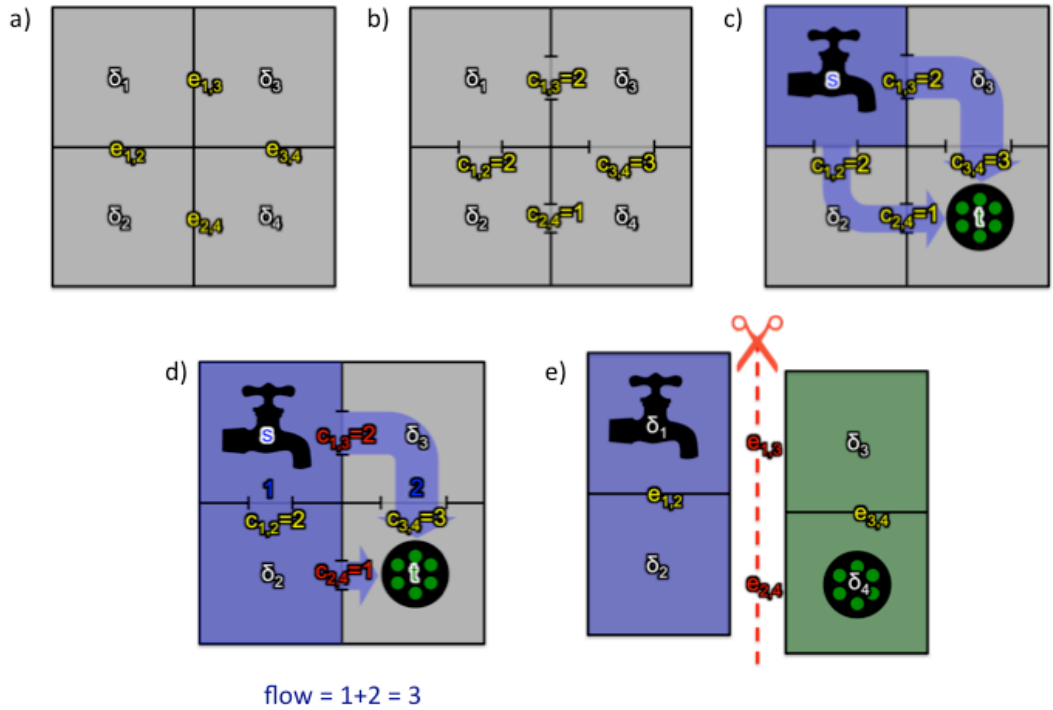

**Figure 2.** Illustration of the max-flow/min-cut theorem adapted after Dantzig and Fulkerson (2003). a) The image is considered as a network consisting of four pixels (nodes) $\delta$ that are connected by four edges $e$. b) Each edge has a capacity $c$. c) Some pixels are considered as source **s** or sink **t**. A flow through the network is envisioned. d) The maximal steady state flow is limited by bottlenecks in the system (indicated in red); the maximal flow is 3 as given by the sum of $c_{1,3}$ and $c_{2,4}$. This process can be understood in terms of the flow analogy: the lower left pixel will saturate at steady state (blue), while the pixels to the right of the bottlenecks (grey area) will not saturate as the water can quickly flow through this part of the system. e) The indicated cut separating the image into two disjoint subsets **S** (area highlighted in blue) and **T** (area highlighted in green) is a set of edges [$e_{1,3}$, $e_{2,4}$]. The cost of the cut is $C = c_{1,3} + c_{2,4} = 2+1 = 3$, equal to the maximal flow from source **s** to sink **t**.

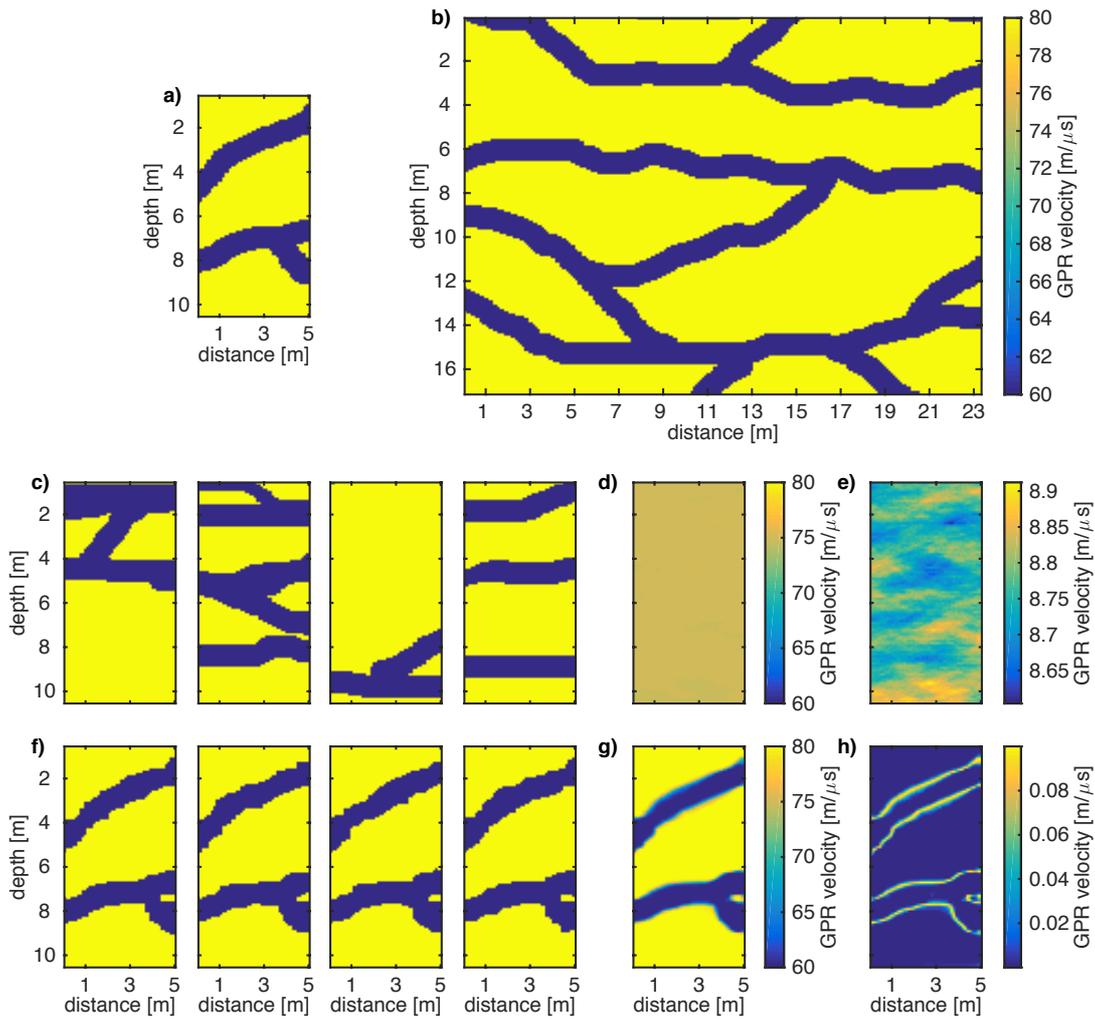

**Figure 3.** Test case I with channel-like structures. a) True velocity field $\mathbf{m}_{\text{ref}}$, b) section of the training image (the size of the entire training image is 250 × 250 m), c) prior model realisations, d) mean of prior model realisations, e) standard deviation of prior model realisations, e) posterior model realisations, f) mean of posterior model realisations, g) standard deviation of posterior model realisations.

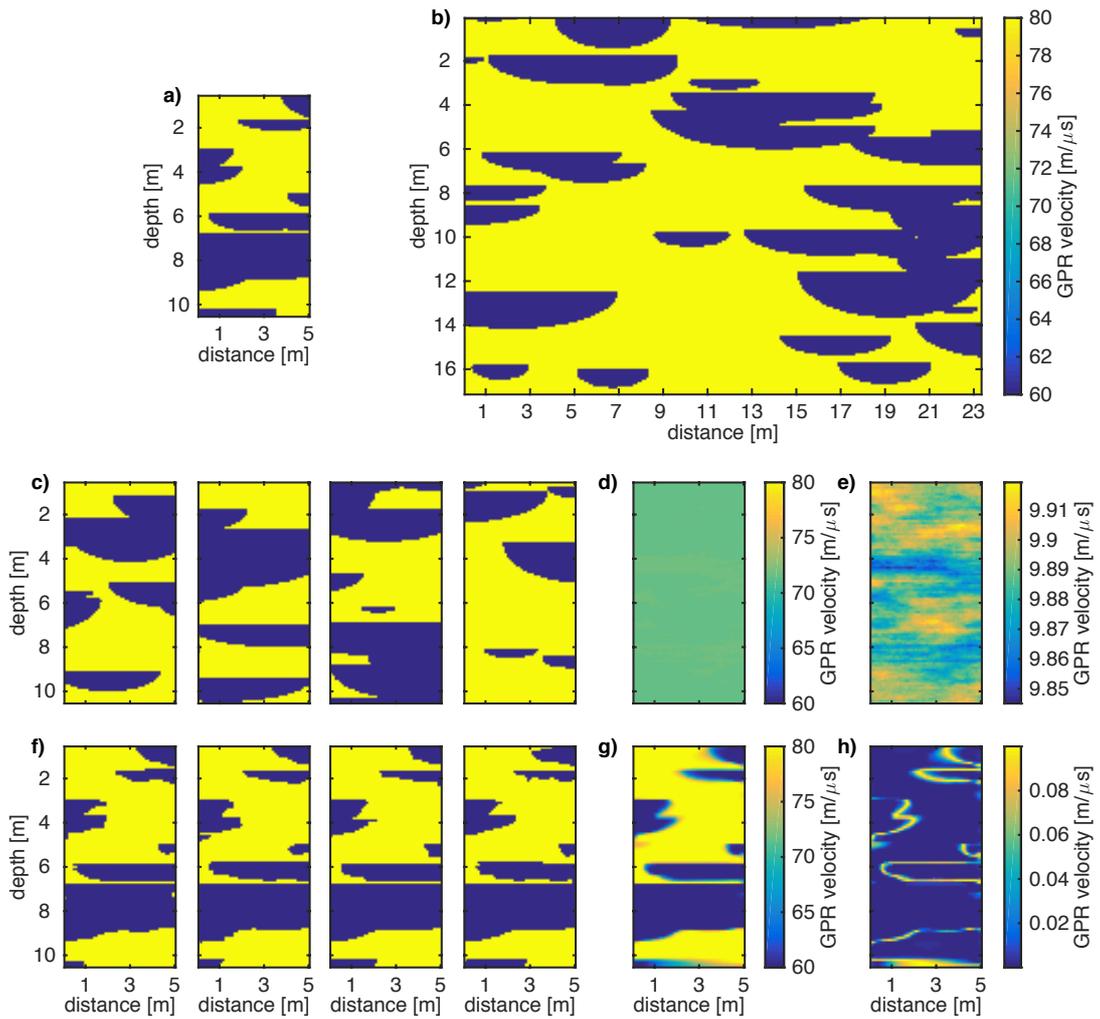

**Figure 4.** Test case II with lens-like structures. a) True velocity field **m**<sub>ref</sub>, b) section of the training image (the size of the entire training image is 250 × 250 m), c) prior model realisations, d) mean of prior model realisations, e) standard deviation of prior model realisations, e) posterior model realisations, f) mean of posterior model realisations, g) standard deviation of posterior model realisations.
32

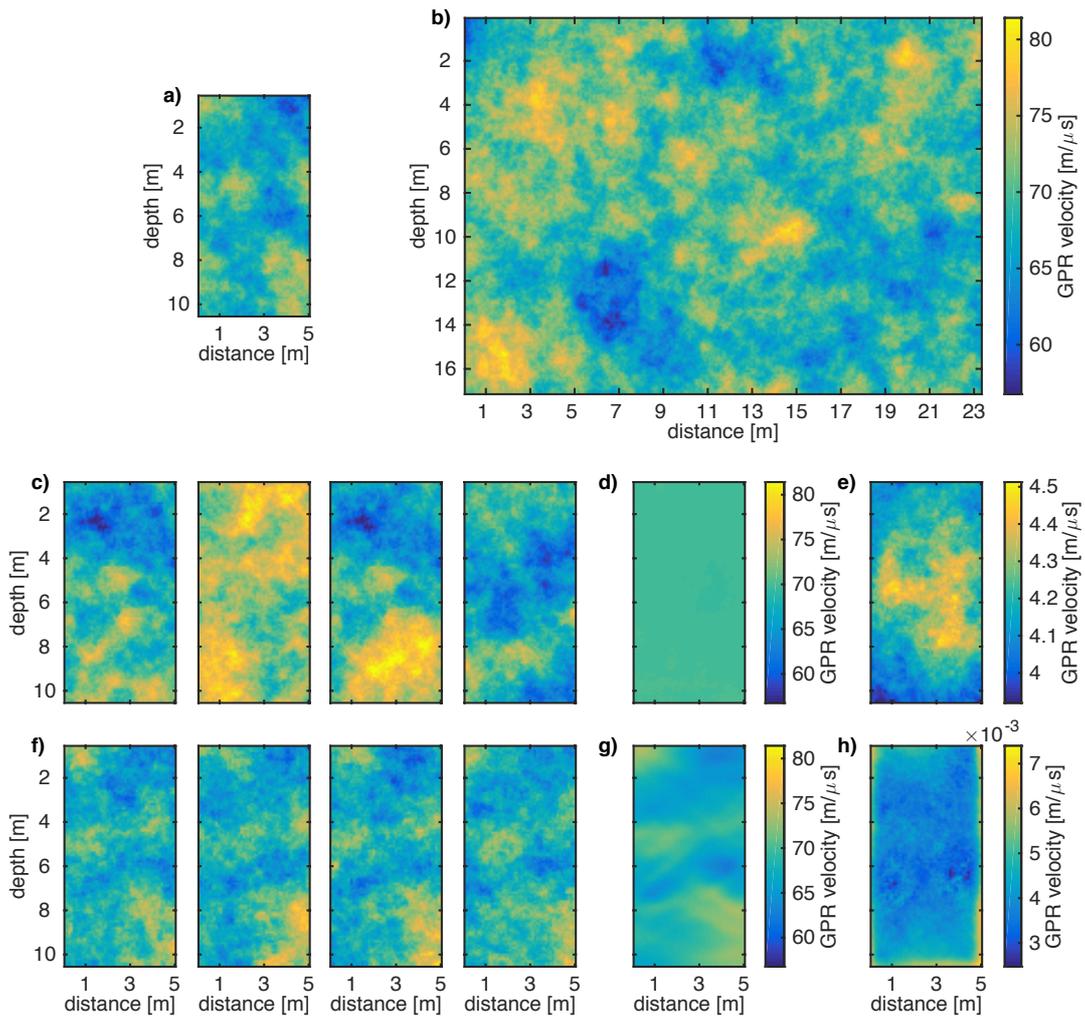

**Figure 5.** Test case III with a multi-Gaussian field. a) True velocity field $\mathbf{m}_{\text{ref}}$, b) section of the training image (the size of the entire training image is 250 × 250 m), c) prior model realisations, d) mean of prior model realisations, e) standard deviation of prior model realisations, e) posterior model realisations, f) mean of posterior model realisations, g) standard deviation of posterior model realisations.

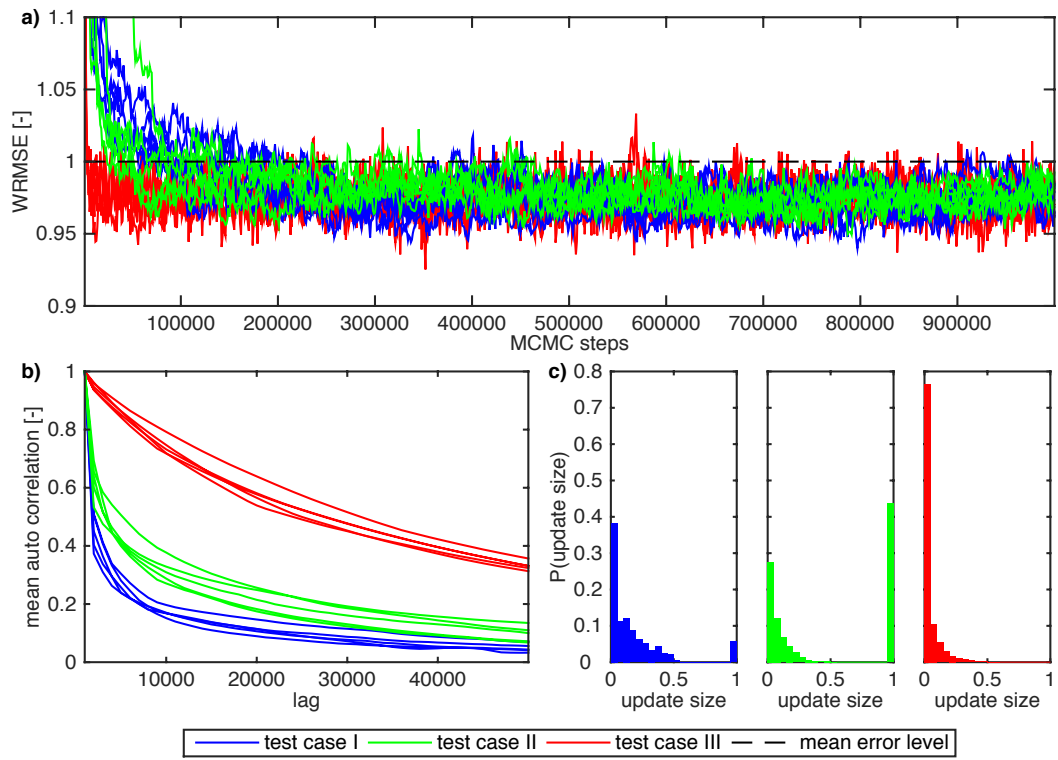

**Figure 6.** a) Data misfit (WRMSE) of model realisations in the MCMC chains, b) mean of the auto-correlation coefficients in the MCMC chains as a function of lag of MCMC steps, c) distribution of the updated model fraction in the model proposals (l.t.r.: test case I, test case II, test case III).

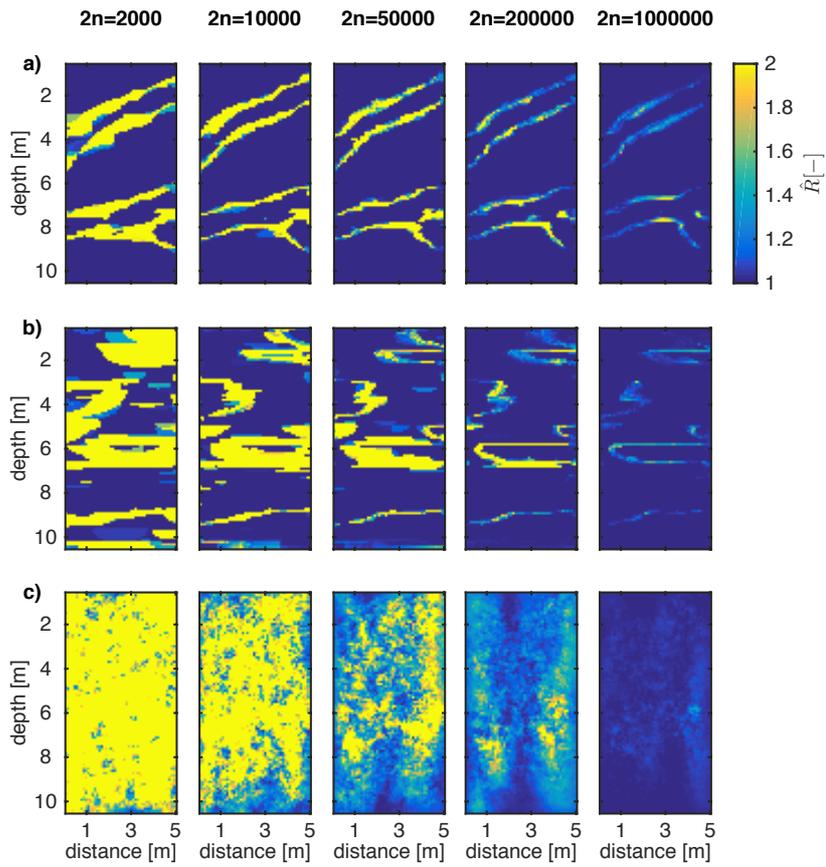

**Figure 7.** A potential scale reduction factor $\hat{R} \leq 1.2$ indicates that the different chains have reached the same limiting distribution. a) Test case I, b) test case II c) test case III. 2*n* is the length of the MCMC chains.

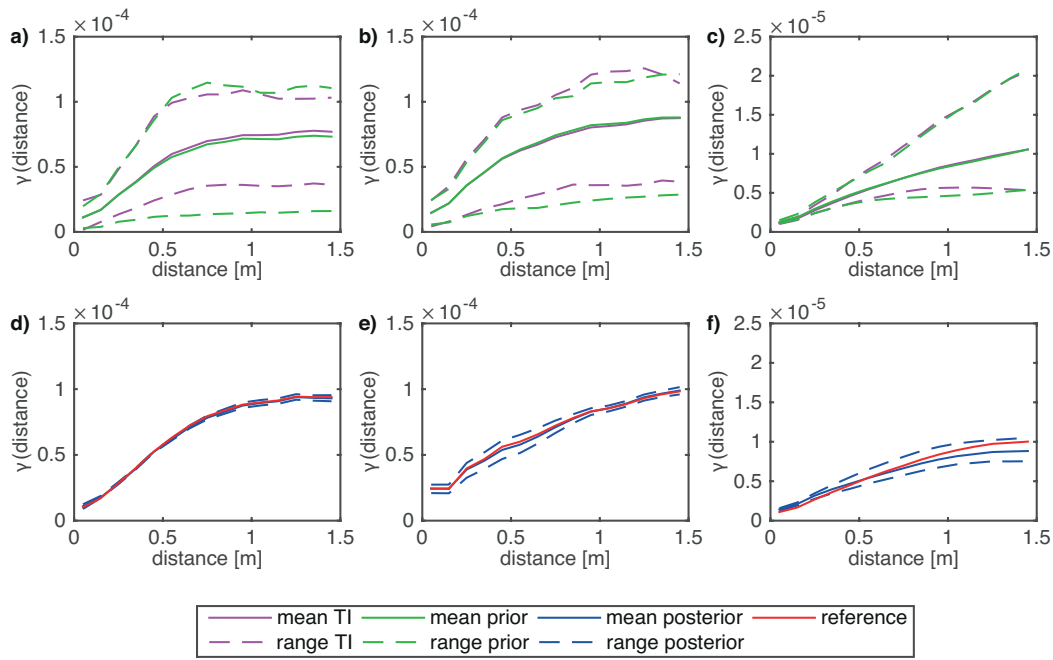

**Figure 8.** a-c) Comparison between the experimental variogram of the training image and the prior realisations for test cases I, II and III, respectively. d-f) Comparison between the experimental variogram of the true model $\mathbf{m}_{\text{ref}}$ and the posterior realisations for test cases I, II and III, respectively.

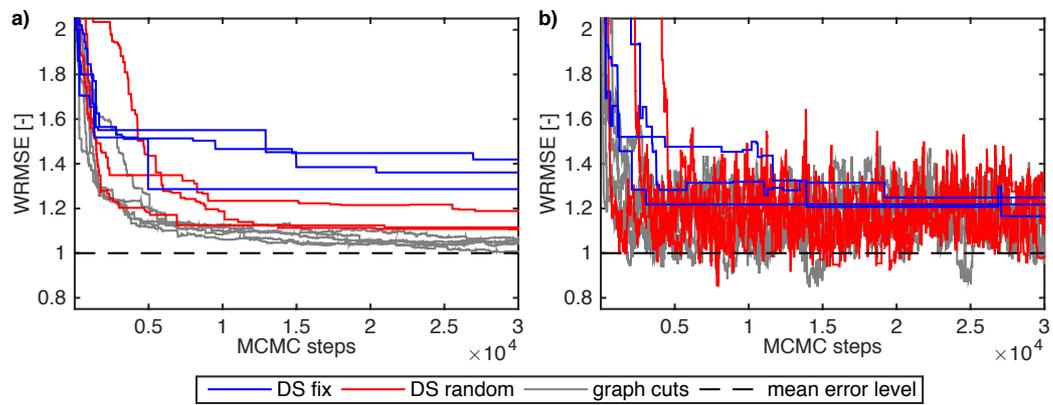

**Figure 9.** Comparison of the burn in phase using model proposals based on graph cuts (our method) and direct sampling. a) Inversion of extensive data *N*=566, b) inversion of sparse data *N*=29.

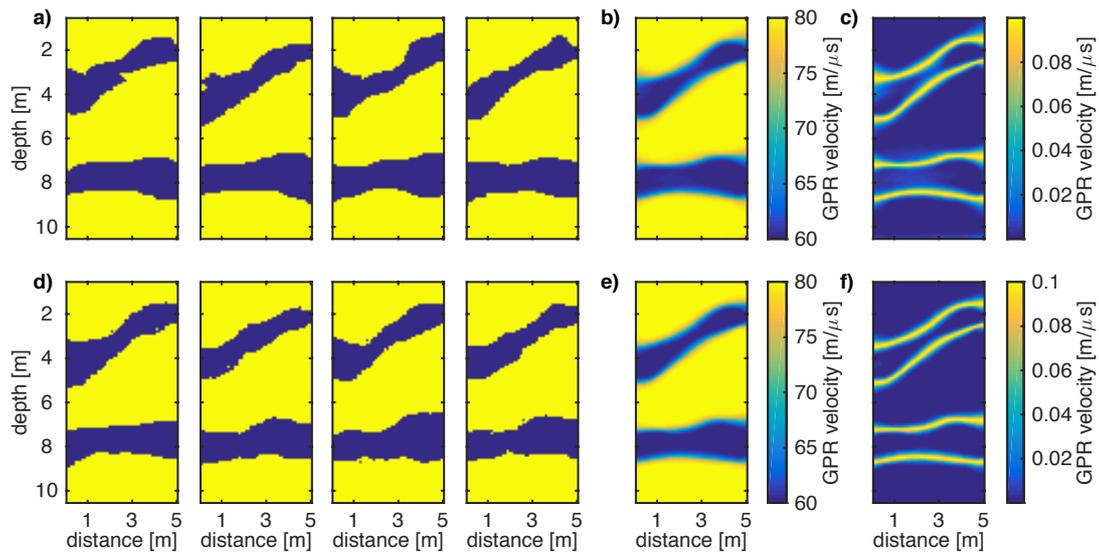

**Figure 10.** Comparison of posterior model realisations created by image synthesis with (a-c) graph cuts and (d-f) direct sampling using sparse data. a) posterior model realisations created by image synthesis, b) mean and c) standard deviation of posterior model realisations created by image synthesis, d) posterior model realisations created by direct sampling, e) mean and f) standard deviation of posterior model realisations created by direct sampling.

**Table 1.** Mean and variance of pixel values of training images.

| Training image | Mean of pixel values | Variance of pixel values |
|---|---|---|
| Test case I | 74.8 m/$\mu$s | 8.8 m/$\mu$s |
| Test case II | 71.5 m/$\mu$s | 9.9 m/$\mu$s |
| Test case III | 70.0 m/$\mu$s | 4.0 m/$\mu$s |